\documentclass[sigconf]{acmart}

\AtBeginDocument{%
  \providecommand\BibTeX{{%
    \normalfont B\kern-0.5em{\scshape i\kern-0.25em b}\kern-0.8em\TeX}}}

\usepackage{booktabs}
\usepackage{tabularx}
\usepackage{array}
\newcolumntype{Y}{>{\raggedright\arraybackslash}X}

\copyrightyear{2026}
\acmYear{2026}
\setcopyright{cc}
\setcctype{by}
\acmConference[RecSys '26]{20th ACM Conference on Recommender Systems}{September 27-October 02, 2026}{Minneapolis, MN, USA}
\acmBooktitle{20th ACM Conference on Recommender Systems (RecSys '26), September 27-October 02, 2026, Minneapolis, MN, USA}
\acmDOI{10.1145/3773078.3841263}
\acmISBN{979-8-4007-2284-4/2026/09}
\begin{document}

\title[Predicting Custom-Feed Returns for New Bluesky Posts: A
Prospective Study]{Predicting Custom-Feed Returns for New Bluesky Posts: A Prospective Study}

\author{Yipeng Wang}
\orcid{0009-0004-3716-8816}
\affiliation{%
  \institution{Northeastern University}
  \city{Boston}
  \state{Massachusetts}
  \country{USA}
}
\email{wang.yipen@northeastern.edu}

\author{Mohit Singhal}
\orcid{0000-0002-7423-9116}
\affiliation{%
  \institution{Northeastern University}
  \city{Boston}
  \state{Massachusetts}
  \country{USA}
}
\email{m.singhal@northeastern.edu}

\renewcommand{\shortauthors}{Yipeng Wang and Mohit Singhal}


\begin{abstract} The conventional approach to cold-start recommendation addresses new users or newly introduced items. Bluesky custom feeds create a different setting: independently operated feeds filter content from a shared public stream. In this setting, newly published posts are the cold-start objects, while the feeds serve as candidates. We propose a cold-start routing task in which a newly ingested public post is the query and all rankable feeds in the monitored panel are ranked according to whether each will subsequently return it. We build a still-evolving collect-first, label-later benchmark dataset. The collected dataset covers a fixed panel of 5,000 monitored feeds and contains 17.804 million public posts, 1.865 million observable post--feed return records, and 625,083 valid feed polls. The labels record whether a post is observed among a feed's AppView Top-50 results in at least one poll during the 24 hours after publication. The current experiments use two disjoint 24-hour test folds, each paired with a 24-hour training window and separated by a 24-hour outcome-availability gap. Evaluation is conditional on the 602,186 test posts that have at least one positive observed label and satisfy the metric eligibility criteria; these posts account for 9.04\% of all 6,661,658 test posts. Across the two folds, LambdaRank achieves the best equal-fold mean values among the evaluated models: 0.7361 for capped Recall@10, 0.6127 for NDCG@10, and 0.7749 for Hit@10. \end{abstract}
\maketitle

\section{Introduction}

Classical cold-start recommendation assumes that, within a fixed system, there are users or items that lack historical behavioral data \cite{schein2002coldstart}, whereas a decentralized feed ecosystem does not fit this assumption. On Bluesky, posts first enter a shared public stream and are then independently filtered by separately operated feed generators, which determine which posts to return \cite{kleppmann2024atproto}. Existing studies have analyzed Bluesky's network structure and algorithmic curation through custom feeds \cite{balduf2024bluesky,quelle2025bluesky}, and have also built personalized feeds for research settings \cite{greenwood2026paperskygest}, but these studies do not evaluate the objective investigated in this paper: predicting whether a post will be returned by a feed in the future.

We name this prediction task cold-start routing: ranking all rankable feeds in a fixed monitored panel based on whether each feed will subsequently return a newly ingested post. We position cold-start routing as a foundational building block for the custom-feed ecosystem, such that the ranking can be used to route each newly ingested post to the feeds likely to return it, and to study how exposure is distributed across the feed marketplace.  Retrospective crawler-based collection can blur the temporal order between post creation and subsequent feed-return behavior. Our dataset first collects public posts and then obtains labels through subsequent feed polls. We describe this collect-first, label-later order of data acquisition as prospective. All models in this paper are evaluated on an offline temporal holdout after all labels have been generated, rather than on real-time replay data from online deployment. This observational use of platform-generated outputs is related to data-driven audits of black-box ranking and recommendation systems~\cite{singhal2025yelp}.
Across the two outcome-purged 24-hour test folds, LambdaRank is the strongest evaluated method, with equal-fold mean capped Recall@10 of 0.7361, NDCG@10 of 0.6127, and Hit@10 of 0.7749.

\section{Task and protocol}

For a public post collected at time $t$, a post--feed pair is labeled positive if the post is observed among the first 50 items returned by the public \path|app.bsky.feed.getFeed| AppView endpoint~\cite{bskyapi2026} within the 24-hour observation window after publication. Data snapshots are collected approximately once per hour to ensure observation timeliness. 
The features include post text/content, feed name and metadata, historically observable author--feed return interaction records, lexical and semantic similarity features, feed labels, and content popularity signals.

This study reports the three core metrics, capped Recall@10, NDCG@10, and Hit@10, only for valid test posts that have at least one positive observed label and satisfy the metric calculation rules. For any post $p$, let $Y_p$ denote its feeds with positive observed labels. Capped Recall@10 is calculated as follows:

\[
\operatorname{cR@10}(p)
=
\frac{\left|\operatorname{Top}_{10}(p)\cap Y_p\right|}
{\min\left(10,\left|Y_p\right|\right)}.
\]

Here, Hit@10 indicates whether any feed in the target feed set $Y_p$ ranks among the top 10 positions. NDCG@10 is calculated using binary relevance weights constructed from the observed return behavior. The two test folds contain 319,908 and 282,278 metric-eligible positive-label posts, respectively, for 602,186 total, accounting for 9.04\% of all 6,661,658 test posts.

\section{Benchmark and Ranking Pipeline}

\textbf{Data collection: } We collected public \path|app.bsky.feed.post| records from the AT Protocol event stream and constructed a fixed panel of 5,000 publicly resolvable custom feeds; feed-return labels were then obtained through the public \path|app.bsky.feed.getFeed| AppView API \cite{kleppmann2024atproto}. 
Posts were collected from June 30 through July 6, 2026 (UTC). This collection includes 17.804 million public posts, 1.865 million observable post--feed returns, and 625,083 valid feed polls. Of the monitored feeds, 3,079 produced at least one observable return record. The fold-specific train/test positive-label proportions are 5.97\%/9.35\% and 6.73\%/8.72\%, respectively.

\textbf{Retrieval and pair representation: } For each query post, we construct six feature-scoring channels to retrieve and score feeds from the fixed 4,233-feed ranking index derived from the 5,000-feed monitored panel. These channels cover historical author--feed match records, lexical TF--IDF similarity, character-level TF--IDF similarity, fixed \path|paraphrase-multilingual-MiniLM-L12-v2| semantic similarity \cite{wang2020minilm,reimers2019sentencebert}, hashtag similarity, and historical match-volume features. The retrieval results from all channels are combined into a candidate set containing at most 370 feeds for subsequent reranking. Feed features integrate pre-origin match records, snapshot text, display names, functional descriptions, and creator account information, providing multidimensional feature representations. Each scored post--feed pair consists of 27 features, covering retrieval scores and ranking positions, historical feed and author behavior, post attributes, and cross-feature interaction information. Four of these features are generated using \path|paraphrase-multilingual-MiniLM-L12-v2| sentence-embedding features, and the encoder uses fixed parameters and does not participate in model fine-tuning.

\begin{figure*}[t]
  \centering
  \includegraphics[width=\textwidth]{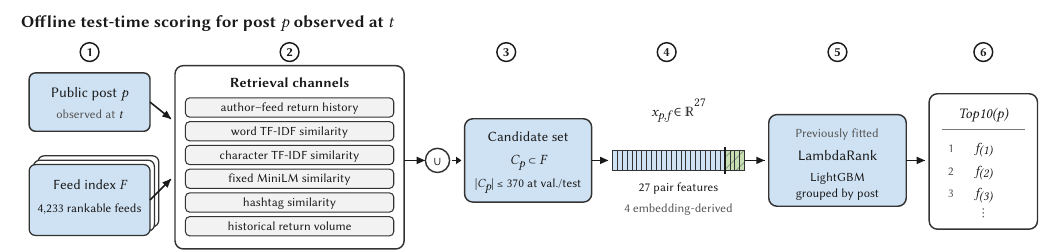}
  \caption{Workflow for cold-start feed routing.
    Given a newly observed post~$p$, six retrieval channels construct
    a candidate union $C_p$ containing at most 370 feeds from the fixed
    5,000-feed panel. The system represents each candidate post--feed pair
    with 27 features, groups candidates by post, and applies a previously
    fitted LambdaRank model to return the top 10 feeds.
    Validation and test outcome labels are excluded from candidate
    generation and scoring.}
    \Description{The main workflow graph}
  \label{fig:workflow}
\end{figure*}

\textbf{Ranking models: } We first establish three single-feature baseline models, which rank the full index based on historical match volume, author--feed match records, and character-level TF--IDF text similarity, respectively. On this basis, different model designs are used to distinguish the scoring mechanism, semantic reranking inputs, and ranking optimization strategy. First, the pointwise histogram gradient boosting classification model (HGB) removes the four semantic embedding features and is trained on candidate-set pair samples using the remaining 23 features. After training, the same HGB model can score and rank either the full index or the candidate set at test time; only the scoring scope changes. Second, the semantically enhanced HGB model follows the same candidate-set training paradigm and additionally introduces the four semantic embedding features, integrating semantic information with traditional statistical features. Finally, the LambdaRank model uses the complete set of 27 features and the same candidate-set samples. During training, candidate pairs are grouped by query post, and the model uses the LambdaRank objective implemented in LightGBM for NDCG optimization~\cite{burges2010lambdamart,ke2017lightgbm}. All learning-based ranking models in this paper use gradient-boosted tree ensemble architectures. The fixed sentence-vector encoder used only for semantic feature extraction is the only neural-network module in the pipeline. Figure~\ref{fig:workflow} summarizes this retrieve-then-rerank process.

\section{Results}
\begin{figure}[h!]
  \centering
  \includegraphics[width=0.82\columnwidth]{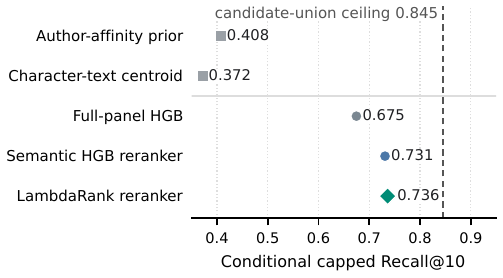}
  \caption{Conditional cR@10; Table~\ref{tab:results} gives the updated fold values and means. The candidate-union ceiling is diagnostic.}
  \Description{A point plot comparing conditional capped Recall at ten for the author-affinity prior, character-level text model, full-panel HGB, semantic HGB, and LambdaRank, with a dashed candidate-union ceiling.}
  \label{fig:recall}
\end{figure}

Table~\ref{tab:results} and Figure~\ref{fig:recall} show the transition from single-signal priors to learned reranking. LambdaRank achieves the best equal-fold mean metrics: 0.7361 capped Recall@10, 0.6127 NDCG@10, and 0.7749 Hit@10; the author-affinity prior is the best-performing single-signal baseline.

\begin{table}[t]
\caption{Two purged 24-hour folds (319,908/282,278 positive-label posts). cR is Fold~1/Fold~2; other columns are equal-fold means.}
\Description{A table of fold-specific and equal-fold capped Recall at ten, equal-fold NDCG at ten, and equal-fold Hit at ten. LambdaRank has the best learned mean values.}
\label{tab:results}
\scriptsize
\setlength{\tabcolsep}{0.65pt}
\renewcommand{\arraystretch}{0.90}
\centering
\begin{tabular}{@{}lcccc@{}}
\toprule
Method & F1/F2 cR & Mean cR & NDCG & Hit \\
\midrule
LambdaRank reranker      & \textbf{.7252/.7471} & \textbf{.7361} & \textbf{.6127} & \textbf{.7749} \\
Semantic HGB reranker    & .7216/.7403 & .7310 & .5972 & .7695 \\
Base HGB reranker        & .6953/.7143 & .7048 & .5751 & .7444 \\
Full-panel HGB           & .6607/.6886 & .6747 & .5523 & .7149 \\
Author-affinity prior    & .3860/.4295 & .4078 & .3636 & .4426 \\
Character TF--IDF        & .3673/.3761 & .3717 & .2468 & .4130 \\
Popularity prior         & .0130/.0170 & .0150 & .0084 & .0217 \\
\midrule
Candidate-union ceiling  & .8382/.8514 & .8448 & -- & .8648 \\
\bottomrule
\end{tabular}
\end{table}

Applying the candidate-trained base HGB to the retrieved union rather than the full ranking index raises mean capped Recall@10 by 0.0301; adding the four semantic reranking features raises it by another 0.0262. Holding the candidate set and 27-feature representation fixed, LambdaRank exceeds semantic HGB by 0.0051 capped Recall@10 and by 0.0155 NDCG@10. Author-clustered bootstrap 95\% CIs are $[0.0043,0.0059]$ (cR) and
$[0.0132,0.0175]$ (NDCG); all cascade gains are positive in all 16
three-hour test blocks.

The equal-fold candidate-union ceiling is 0.8448 capped Recall@10 and contains at least one observed positive for 86.48\% of positive-label posts. Its 0.1552 deficit from the ideal value is larger than the 0.1087 gap from the ceiling to LambdaRank, identifying candidate retrieval as the larger remaining source of metric loss. Across all 6,661,658 test arrivals, 466,286 (7.00\%) receive a LambdaRank Top-10 observed-positive hit.

\section{Limitations and Future Work}
The collected post--feed pairs are only a small subset of the Bluesky universe. More specifically, we monitored only the Top-50 posts at each polling time for 5,000 feeds for only 1 week. Returns outside the panel or below rank 50 are unobserved, so unobserved pairs cannot be interpreted as confirmed negatives. The two consecutive daily test folds do not establish broad cross-day generality. Future work could expand the collection period and coverage and target platform-wide inclusion, private activity, impressions, or user responses.

\section{Conclusion}

We introduce cold-start feed routing and a collect-first, label-later benchmark. Across two outcome-purged daily folds, LambdaRank leads, while candidate-ceiling analysis identifies retrieval as the larger remaining deficit.

\bibliographystyle{ACM-Reference-Format}
\bibliography{references}

\end{document}